\documentclass[aps,prl,twocolumn,superscriptaddress]{revtex4-2}

\usepackage{graphicx}
\usepackage{color}
\usepackage{amsmath,amssymb}
\usepackage{soul}
\newcommand{\ket}[1]{|#1\rangle}
\newcommand{\bra}[1]{\langle#1|}

\newcommand{\wc}[1]{\textcolor{black}{#1}}
\newcommand{\wcr}[1]{\textcolor{black}{#1}}
\newcommand{\wcrr}[1]{\textcolor{black}{#1}}
\newcommand{\krev}[1]{\textcolor{black}{#1}}

\begin{document}
\setlength{\pdfpagewidth}{8.5in}
\setlength{\pdfpageheight}{11in}


\title{Quantum jumps in the non-Hermitian dynamics of a superconducting qubit}

\author{Weijian Chen}\email{wchen34@wustl.edu}
\affiliation{Department of Physics, Washington University, St. Louis, MO, USA}
\affiliation{Center for Quantum Sensors, Washington University, St. Louis, MO, USA}
\author{Maryam Abbasi}
\affiliation{Department of Physics, Washington University, St. Louis, MO, USA}
\author{Yogesh N. Joglekar}
\affiliation{Department of Physics, Indiana University-Purdue University Indianapolis, Indianapolis, IN, USA}
\author{Kater W. Murch}\email{murch@physics.wustl.edu}
\affiliation{Department of Physics, Washington University, St. Louis, MO, USA}
\affiliation{Center for Quantum Sensors, Washington University, St. Louis, MO, USA}

\date{\today}

\begin{abstract}

We study the dynamics of a \wcr{driven} non-Hermitian superconducting qubit which is perturbed by quantum jumps between energy levels, a purely quantum effect with no classical correspondence. The quantum jumps mix the qubit states leading to decoherence. We observe that this decoherence rate is enhanced near the exceptional point, owing to the cube-root topology of the non-Hermitian eigenenergies. Together with the effect of non-Hermitian gain/loss, quantum jumps can also lead to a breakdown of adiabatic evolution under the slow-driving limit. Our study shows the critical role of quantum jumps in generalizing the applications of classical non-Hermitian systems to open quantum systems for sensing and control.

\end{abstract}

\maketitle

Dissipation is ubiquitous in nature; as in 
radioactive decay of an atomic nucleus and wave propagation in absorptive media, dissipation results from the coupling of these systems to different environmental degrees of freedom. These dissipative systems can be phenomenologically described by effective non-Hermitian Hamiltonians, where the non-Hermitian terms are introduced to account for the dissipation. The non-Hermiticity leads to a complex energy spectrum with the imaginary part quantifying the loss of particles/energy from the system. The degeneracies of a non-Hermitian Hamiltonian are known as exceptional points (EPs), where both the eigenvalues and the associated eigenstates coalesce \cite{Miri2019, Ozdemir2019}. The existence of EPs has been demonstrated in many classical systems \cite{Dembowski2001,Ruter2010, Peng2014a,Schindler2012,Bender2013,Shi2016, Zhu2014, Li2019,Partanen2019}
with applications in laser mode management \cite{Peng2016, Brandstetter2014, Wong2016}, enhanced sensing \wcr{\cite{Wiersig2014, Chen2017, Hodaei2017, Langb2018, Lau2018, Zhang2019}}, and topological mode transfer \cite{Xu2016, Doppler2016,Choi2017,Zhang2018}. 

\begin{figure}
    \centering
    \includegraphics{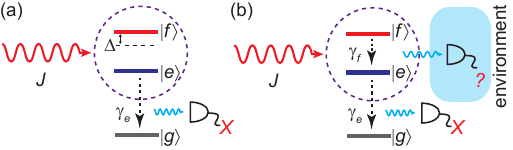}
    \caption{(a) Formation of a non-Hermitian qubit through a dissipative three-level system. The ground level $|g\rangle$ acts as a continuum and can be used to monitor the quantum jumps from the \{$|e\rangle, |f\rangle$\} submanifold. When postselection is used to eliminate this dynamics, the evolution in the \{$|e\rangle, |f\rangle$\} submanifold is \wcr{governed by a non-Hermitian Hamiltonian}. $J$ denotes the coupling rate from an applied drive with frequency detuning $\Delta$ relative to the $|e\rangle$---$|f\rangle$ transition, and $\gamma_e$ denotes the dissipation rate of the $|e\rangle$ level.
    (b) The quantum jumps from the $|f\rangle$ level at rate $\gamma_f$ are only recorded by the environment, and this missing information \wcr{necessitates a hybrid-Liouvillian formalism (see main text).}
    }
    \label{fig1}
\end{figure}

\wcrr{Though the effective Hamiltonian approach has been developed decades ago as part of quantum measurement theory,}
recent experiments with single electronic spins \cite{Wu2019,liu2020}, superconducting qubits \cite{Naghiloo2019}, and photons \cite{Xiao2017,Klauck2019,Yu2020}  
have \wcrr{expanded interest in} uniquely quantum effects in non-Hermitian dynamics. Two approaches have been taken to study non-Hermitian dynamics in the quantum regime. The first is to simulate these dynamics---through a process known as Hamiltonian dilation---by embedding a non-Hermitian  Hamiltonian into a larger Hermitian system \cite{Wu2019,liu2020,Yu2020}. A second approach is to  directly isolate the non-Hermitian dynamics from a dissipative quantum system \cite{Naghiloo2019}. To understand this approach, recall that dissipative quantum systems are usually described by a Lindblad master equation that contains two dissipative terms: the first is a term that describes quantum jumps between the energy eigenstates of the system, and the second is a term that yields coherent non-unitary evolution \cite{Dalibard1992, Mlmer1993, Plenio1998}. By suppressing the former term, the resulting evolution is described by an effective non-Hermitian Hamiltonian. This can be achieved through post-selection to eliminate trajectories that contain quantum jumps (Fig.~\ref{fig1}(a)) \cite{Naghiloo2019}. However, additional sources of \wcr{energy dissipation or pure dephasing }
can alter this non-Hermitian evolution \wcr{\cite{Minganti2019,Minganti2020}}. The combination of non-unitary dynamics and decoherence will lead to evolution that is starkly different than what is encountered with conventional \wcr{open quantum} 
systems.  In this letter, we characterize these dynamics using experiments on a superconducting qutrit. We observe quantum dynamics that result from the competition of the non-unitary effect of complex energies and quantum jumps. This leads to decoherence enhancement near the EP, non-stationary evolution of system eigenstates, and a quantum jump-induced breakdown of adiabaticity \wcr{when a system parameter is slowly varied}. 

Our experiment uses the lowest three energy levels ($\vert g \rangle$, $\vert e \rangle$ and $\vert f \rangle$) of a transmon superconducting circuit \cite{Koch2007} that consists of a pair of Josephson junctions in a SQUID geometry shunted by a capacitor. The transmon circuit is placed within a three-dimensional copper microwave cavity that serves two purposes in the experiment. First, it mediates the interaction between the circuit and a \wcr{nonuniform} density of states of the electromagnetic field, allowing us to tune the dissipation rates of the transmon energy levels such that $\gamma_e$ (the decay rate of the $|e\rangle$ level) is much larger than $\gamma_f$ (the decay rate of the $|f\rangle$ level). Second, the dispersive interaction between cavity mode and the circuit results in a state-dependent cavity resonance frequency \cite{Wallraff2005}. We achieve high-fidelity, single-shot readout of the transmon state by probing the cavity with a weak microwave signal and detecting its phase shift.



The dynamics of this three-level quantum system (Fig. \ref{fig1}(b)) is described by a  Lindblad master equation 
\begin{equation}
\frac{\partial \rho_{\mathrm{tot}}}{\partial t} = -i [H_\mathrm{c}, \rho_{\mathrm{tot}}] + \sum_{k=e,f} [ L_k \rho_{\mathrm{tot}} L_k^\dag - \frac{1}{2} \{L_k^\dag L_k, \rho_{\mathrm{tot}}\}],
\label{eq:Lindblad}
\end{equation}
where $\rho_{\mathrm{tot}}$ denotes a $3 \times 3$ density operator. The 
\wcr{jump operators} $L_e = \sqrt{\gamma_e} |g\rangle \langle e|$ and $L_f = \sqrt{\gamma_f} |e\rangle \langle f|$ describe the energy decay from $|e \rangle$ to $|g \rangle$ and from $|f \rangle$ to $|e \rangle$, respectively. Here we only consider a drive at the \{$|e\rangle$, $|f\rangle$\} submanifold, and in the rotating frame $H_\mathrm{c} = J (|e \rangle \langle f| + |f \rangle \langle e|) + \Delta/2 (|e \rangle \langle e| - |f \rangle \langle f|)$, where $\Delta$ is the frequency detuning (relative to the $|e\rangle$---$|f\rangle$ transition) of the microwave drive that couples the states at rate $J$. \wcr{The dissipative dynamics of the qutrit to its steady state can be captured by a Liouvillian superoperator defined by Eq. (\ref{eq:Lindblad}), exhibiting the so-called Liouvillian EPs \cite{Hatano2019,Minganti2019,Arkhipov2020_02}. A classical analogy of this EP transition is a damped harmonic oscillator, where an EP (corresponding to the critical damping) marks the transition from overdamped to underdamped regime \cite{Hatano2019,Minganti2019,Arkhipov2020_02}. 
}

We utilize the high fidelity single shot readout to isolate 
dynamics in the $\{\ket{e},\ket{f}\}$ submanifold by eliminating any experimental trials where the qubit undergoes a jump to the state $\ket{g}$  \cite{Naghiloo2019}.  The resulting dynamics in the submanifold is governed by
\begin{equation}
\frac{\partial \rho}{\partial t} = -i (H_\mathrm{eff}\rho - \rho H_\mathrm{eff}^{\dagger})+L_f \rho L_f^\dag 
\label{eq:hybridLindblad}
\end{equation}
where $\rho$ denotes a $2 \times 2$ density operator. The effective non-Hermitian Hamiltonian $H_\mathrm{eff} = H_\mathrm{c} -iL_e^\dag L_e/2 -iL_f^\dag L_f/2$ takes into account the coherent nonunitary dissipations of both levels and possesses a second-order EP at $J_\mathrm{EP} = (\gamma_e - \gamma_f)/4$ and $\Delta=0$. This EP separates \wcr{``broken'' and ``unbroken'' regions of effective parity-time ($\mathcal{PT}$) symmetry}, where the difference between eigenvalues is either purely imaginary, or purely real. As shown in Eq.~(\ref{eq:hybridLindblad}), if there are no quantum jumps from the $\ket{f}$ level ($L_f=0$), \wcr{or these jumps can be removed from the dynamics using post-selection}, the system would evolve coherently under $H_\mathrm{eff}$.

\wcr{To capture the effect of jumps from the $|f\rangle$ level, we adopt a hybrid-Liouvillian formalism \cite{Minganti2020}, which describes the non-Hermitian dynamics of an open quantum system under different post-selection efficiencies.}
The dissipative dynamics of the qubit is then written as, 
\begin{equation}
\frac{\partial \rho}{\partial t} = (\mathcal{L}_0+\mathcal{L}_1) \rho. 
\label{eq:liouvilliandiffq}
\end{equation}
Here, the qubit dynamics is captured by two \wcr{hybrid-}Liouvillian superoperators $\mathcal{L}_0 \rho \equiv -i (H_\mathrm{eff} \rho - \rho H_\mathrm{eff}^{\dagger})$ and $\mathcal{L}_1 \rho \equiv L_f \rho L_f^\dag$. \wcr{Compared to the Liouvillian superoperator defined by Eq.~(\ref{eq:Lindblad}), the hybrid-Liouvillian superoperator does not lead to a completely positive and trace-preserving map \cite{Minganti2020}.} \wcr{In contrast to the non-Hermitian Hamiltonian approach based on a Hilbert space of dimension $N=2$, this hybrid-Liouvillian formalism is based on a Liouville space of dimension $N^2=4$.} In the Liouville space, $\rho$ is represented as a $4\times1$ 
vector, and $\mathcal{L}_{i=0,1}$ is represented as a $4\times4$ %
non-Hermitian matrix. Because $\mathcal{L}_0$ encodes the evolution due to $H_\mathrm{eff}$, it also exhibits an EP \wcr{(denoted as `hLEP')} at $J=(\gamma_e-\gamma_f)/4$ and $\Delta=0$. 
One key difference is that three eigenvectors of $\mathcal{L}_0$ coalesce at the EP, implying that a second-order Hamiltonian EP corresponds to a third-order \wcr{hLEP} \cite{Wiersig2020}. 

\wcr{In addition, the non-Hermitian qubit can also suffer from pure dephasing, described by a jump operator $L_\phi = \sqrt{\gamma_\phi/2} \sigma_z$. Its effect includes two aspects: on one hand, it modifies  $H_\mathrm{eff}$ (and subsequently $\mathcal{L}_0$) by adding a term $-i\gamma_\phi I/4$ ($I$ denotes an identity operator), which only shifts the overall loss and does not affect the position of EP in the parameter space; on the other hand, it provides another perturbation of quantum jumps
, the effect of which can be included in $\mathcal{L}_1$ \cite{supp}.}

\wcr{Before proceeding to our experiments, we summarize the possible scenarios of non-Hermitian dynamics of an open quantum system. Depending on the post-selection efficiency $\eta$ of quantum jumps through all possible channels, the resulting non-Hermitian dynamics is described by: (i) a Liouvillian superoperator with $\eta=0$, i.e., no post-selection; (ii) a hybrid-Liouvillian superoperator with $0< \eta <1$, i.e., imperfect post-selection; (iii) an effective non-Hermitian Hamiltonian with $\eta=1$, i.e., perfect post-selection \cite{Minganti2020}. In the semiclassical limit, the quantum jumps are neglected, that is, equivalent to $\eta=1$; subsequently, the dynamics of a dissipative classical system can be described by an effective non-Hermitian Hamiltonian \cite{Minganti2019}.
}

\begin{figure}
    \centering
    \includegraphics{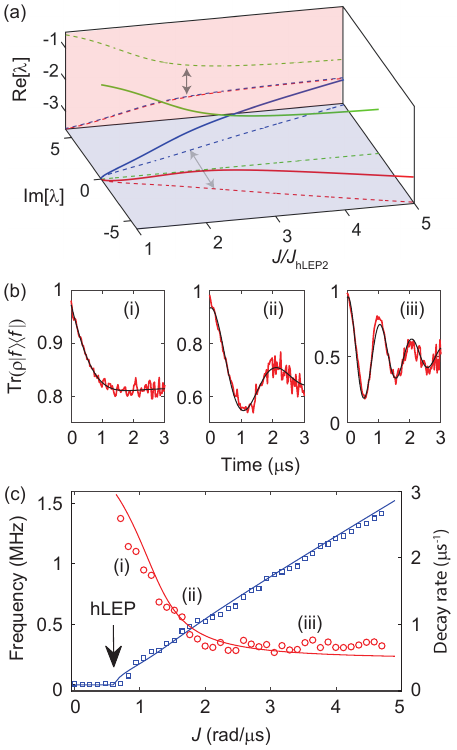}
    \caption{(a) Complex eigenvalues of the \wcr{hybrid-}Liouvillian superoperator $\mathcal{L}_0+\mathcal{L}_1$ in the unbroken regime (solid curves). The dashed curves are the projections of the eigenvalues on the $J$---$\mathrm{Re}[\lambda]$ and $J$---$\mathrm{Im}[\lambda]$ planes. The arrows mark the eigenvalue difference. $J$ is normalized by the value at the second-order \wcr{hybrid-}Liouvillian EP \wc{($J_\mathrm{hLEP2}$)}. Only three of the four Liouvillian eigenvalues involved in this study are shown. \wcr{The blue dashed curve on the $J$---$\mathrm{Re}[\lambda]$ plane has been slightly offset for clarity.} (b) Population dynamics versus evolution time for three different values of $J$, marked by (i)-(iii) in (c). The red curves are experimental results, and the black curves are fits to decaying sine function. (c) The measured oscillation frequency (blue squares, left axis) and decay rate (red circles, right axis) for different drive amplitudes $J$. The solid lines are calculated from the Liouvillian spectra, 
    where the dissipation rates $\gamma_e = 4.5\,\mathrm{\mu s^{-1}}$, $\gamma_f = 0.3\,\mathrm{\mu s^{-1}}$ and $\gamma_\phi = 0.5\,\mathrm{\mu s^{-1}}$ are used.  
    }
    \label{fig2}
\end{figure}

\begin{figure}
    \centering
   \includegraphics{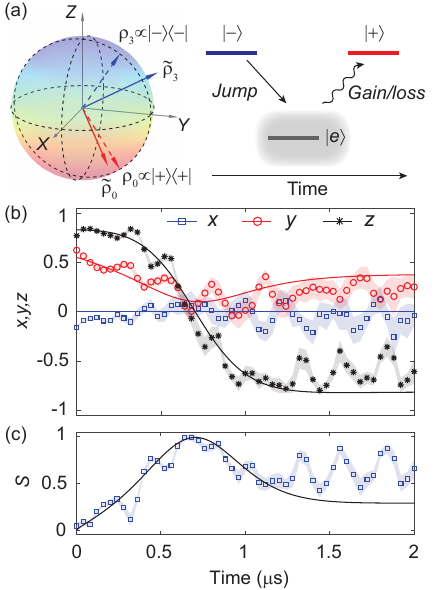}
    \caption{(a) Left: Illustration of the eigenstates and Liouvillian eigenvectors on the Bloch sphere. With no quantum jumps, the two eigenstates \wcr{$|\pm\rangle$} of $H_\mathrm{eff}$ represent the same states as two of the Liouvillian eigenvectors (\wcr{$\rho_{0,3}$,} dashed blue and red arrows). With quantum jumps, the two eigenvectors are perturbed: one corresponds to the steady state (\wcr{$\tilde{\rho}_0$,} solid red arrow); the other one is \wcr{outside of the Bloch sphere and corresponds to} an unphysical state (\wcr{$\tilde{\rho}_3$,} blue solid arrow). Right: Illustration of one quantum trajectory with the qubit prepared at $|-\rangle$, where the qubit first jumps to the $|e\rangle$ level and then evolves to $|+\rangle$ due to the non-Hermitian gain/loss effect. (b, c) Time evolution of the Bloch components (b) and the entropy (c) with the qubit initially prepared at the eigenstate of $H_\mathrm{eff}$ with more loss. $J=0.85\,\mathrm{rad\,\mu s^{-1}}$ places the system in the $\mathcal{PT}$-symmetry broken regime. The symbols are experimental results, \krev{with shaded bands indicating the experimental uncertainty \cite{supp},} and the curves are theoretical results from Eq.~\ref{eq:liouvilliandiffq}. Parameters used are: $\gamma_e = 6.38\,\mathrm{\mu s^{-1}}$, $\gamma_f = 0.24\,\mathrm{\mu s^{-1}}$, $\gamma_\phi = 0.9\,\mathrm{\mu s^{-1}}$. \wcr{Residual oscillations in the data are likely due to technical fluctuations of the tomography calibration.}}
    \label{fig3}
\end{figure}

We first investigate the quantum-jump-induced decoherence in the $\mathcal{PT}$-symmetry unbroken regime.  Figure \ref{fig2}(a) shows the complex eigenvalues \wcr{$\lambda$} of the \wcr{hybrid-}Liouvillian superoperator ($\mathcal{L}_0 + \mathcal{L}_1$) in the unbroken regime, with the real and imaginary parts indicated as projections. Note that the role of the imaginary/real parts of Hamiltonian and \wcr{hybrid-}Liouvillian spectra are reversed because the  `$-i$' term in Eq.~\ref{eq:hybridLindblad} is absorbed into the \wcr{hybrid-}Liouvillian superoperator. The perturbative effect of quantum jumps ($\mathcal{L}_1$) lifts the degeneracy of the third-order \wcr{hLEP} 
of $\mathcal{L}_0$ and generates a new second-order \wcr{hLEP} 
(see \cite{supp} for further details). By lifting the degeneracy, this perturbation leads to decoherence, whose rate is determined by the real part of the eigenvalue difference. The effect of the perturbation is enhanced by proximity to the EP due to the cube-root topology of the third-order degeneracy of $\mathcal{L}_0$ \wcr{\cite{Minganti2020,supp}}.


To experimentally measure the decoherence rates in the vicinity of the \wcr{hL}EP, we initialize the circuit in the $|f\rangle$ state and then apply a microwave drive with amplitude $J$. \wcr{We take $10^4$ measurements per time point and only keep the results with the transmon remaining in the $\{\ket{e},\ket{f}\}$ submanifold for analysis.} We record the final $|f\rangle$ population as a function of time. These dynamics are characterized by damped oscillatory behavior of the population as shown in Fig.~\ref{fig2}(b). We extract the decoherence rate and oscillation frequency for different values of $J$ as shown in Fig.~\ref{fig2}(c). The observed damping rates and oscillation frequencies are in good agreement with the real and imaginary parts of \wcr{hybrid-}Liouvillian spectra, respectively.   In particular, we note that by proximity to the \wcr{hL}EP, the dissipation is dramatically enhanced over its background rate \wcr{(i.e., the rate when far from the hLEP).} 

We now turn to the $\mathcal{PT}$-symmetry broken regime, where \wcr{the quantum jumps compete with the relative non-Hermitian gain/loss effect.}
In the absence of quantum jumps, the qubit has two stationary states, corresponding to the two eigenstates $|\pm\rangle$ of $H_\mathrm{eff}$ (Fig. \ref{fig3}(a)). The corresponding eigenvalues are purely imaginary. Recalling that imaginary eigenvalues correspond to gain or loss, here with $0\geq \mathrm{Im}[\lambda_+] > \mathrm{Im}[\lambda_-]$, both states exhibit loss but the $\ket{+}$ state has gain relative to $\ket{-}$. Therefore, the non-Hermitian dynamics favor the $\ket{+}$ state. 
The 
eigenmatrices $\rho_{0,3}$ of $\mathcal{L}_0$ with the \wcr{smallest} and largest damping rates represent the same states, i.e., $\rho_0 \propto |+\rangle \langle+|$, $\rho_3 \propto |-\rangle \langle -|$. 
\wcr{With the perturbation of quantum jumps ($\mathcal{L}_1$),} the eigenmatrix \wcr{$\rho_0$} becomes slightly mixed \wcr{$\rho_0\to \tilde{\rho}_0$} and corresponds to the \wcr{effective} steady state \wcr{of the non-Hermitian qubit}, while the eigenmatrix \wcr{$\rho_3\to\tilde{\rho}_3$}, a state that is not physically accessible (Fig.~\ref{fig3}(a)). 

\wcr{The physical intuition can be understood at the quantum trajectory level. Given a trajectory with only a single quantum jump from the $|f\rangle$ to $|e\rangle$ level, the jump places the qubit in a superposition of $|+\rangle$ and $|-\rangle$. After that, though no further quantum jumps occur, 
the non-Hermiticity (of $H_\mathrm{eff}$) selects the eigenstate with less loss, and the renormalization of the state subsequently leads to a non-exponential decay \cite{Dalibard1992, Mlmer1993, Plenio1998}. Hence, the eigenstate $|-\rangle$ of $H_\mathrm{eff}$ is unstable and will decay to a steady state in a process that involves both quantum jumps and the non-Hermitian (gain/loss) evolution. Figure~\ref{fig3}(a) displays an illustration of one possible trajectory. Since the trajectories contain unknown number of quantum jumps, the steady state is slightly mixed rather than at the eigenstate $|+\rangle$.}

This prediction is experimentally confirmed through quantum state tomography. Here, we prepare the qubit at the eigenstate $\ket{-}$ of $H_\mathrm{eff}$, and measure the expectation values of the qubit Pauli operators; $\{x,y,z\}\equiv \{\langle \sigma_x\rangle, \langle \sigma_y\rangle,\langle \sigma_z\rangle\}$. Figure~\ref{fig3}(b) displays these expectation values for different evolution times. We highlight several features of the evolution that are different than the dissipative evolution of a Hermitian qubit, where we expect exponential decay to steady state.  The non-Hermitian evolution, perturbed by quantum jumps exhibits  \emph{i)} non-exponential evolution, \emph{ii)} occuring on a  timescale much faster than the quantum jump rate $\gamma_f$.  This occurs due to the non-zero overlap $\langle -|f\rangle$; jumps from $\ket{f}$ to $\ket{e}$ create a mixed state. Thereafter the relative gain of the $\ket{+}$ state causes its population to grow, leading to the non-exponential population evolution. This is further confirmed by examining the evolution of the entropy, defined as $S \equiv - \sum{p_i \log_2 (p_i)}$, where $p_i$ is the eigenvalue of the density matrix $\rho$ of the qubit (Fig. \ref{fig3}(c)). The quantum jumps increase the entropy; this distinguishes the evolution from imperfect (pure) eigenstate preparation, which would also seed non-Hermitian evolution toward $\ket{+}$, but with fixed zero entropy \wc{\cite{Brody2012}}.

\begin{figure}
    \includegraphics{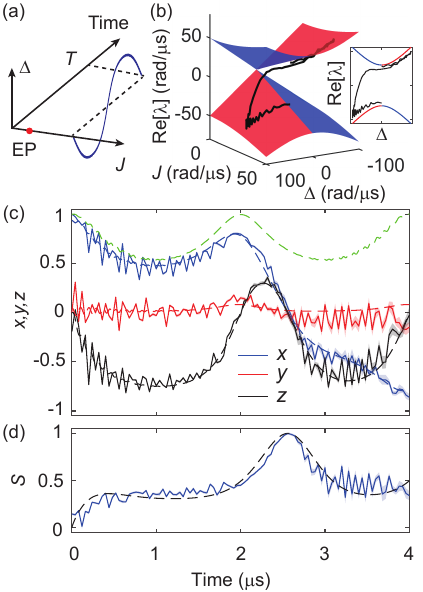}%
    \caption{\label{fig4} (a) Illustration of the parameter path in the parameter space. (b) The real part of the Riemann surface, where the red (blue) surface represents the energy surface  with relative gain (loss). The real part of the energy of the qubit ($\mathrm{Tr}[\rho H_\mathrm{eff}]$) along the trajectory is plotted on the Riemann surface (black line) \wcr{and also shown in the inset}. The time evolution of the Bloch components (c) and the entropy of the corresponding density matrix (d) are displayed for the initial state $|\!+\!x\rangle$ and the loop period $T=4\,\mathrm{\mu s}$. The solid curves are experimental results, \wcrr{with shaded bands indicating the experimental uncertainty \cite{supp},} and the dashed curves are calculations from Eq.~\ref{eq:liouvilliandiffq}. For comparison, the evolution of the $x$ component with no quantum jumps (due to $\mathcal{L}_0$ only) is also shown (dashed green curve in (c)). Parameters used are: $\gamma_e = 6.34\,\mathrm{\mu s^{-1}}$, $\gamma_f = 0.26\,\mathrm{\mu s^{-1}}$, $\gamma_\phi = 0.5\,\mathrm{\mu s^{-1}}$.
    }
\end{figure}

Finally, we study the qubit dynamics under slow parameter variation to reveal the effects of quantum jumps on non-Hermitian adiabatic evolution. We choose a straight parameter path with $J=30\, \mathrm{rad\,\mu s^{-1}}$ ($\gg J_\mathrm{EP} = 1.5\, \mathrm{rad\,\mu s^{-1}}
$) and $\Delta = -30\pi \sin (2\pi t/T)\, \mathrm{rad\,\mu s^{-1}}$, where $T=4\,\mathrm{\mu s}$ is the loop period (Fig. \ref{fig4}(a)). The initial state at $t=0$ is chosen to be an eigenstate of $H_\mathrm{eff}$ (approximated as $|\!+\!x \rangle$). Along this parameter path, the energy gap is large enough to satisfy the slow-driving condition $T|\lambda_+ - \lambda_- | \gg 1$. For $t<T/2$, the initial state follows the instantaneous eigenstate $\ket{+}$ with relative gain.  At $t=T/2$, the parameter path crosses a branch cut for the imaginary Riemann surface at $\Delta=0$. Here, the instantaneous eigenstates exhibit a loss-switch behavior; the eigenstate with relative gain becomes the eigenstate with more loss (Fig.~\ref{fig4}(b)).

The results of quantum state tomography are shown in Fig.~\ref{fig4}(c). At $t=2\,\mathrm{\mu s}$, adiabatic evolution would return the qubit to the state $|\!+\!x \rangle$. The qubit returns close to this state, with slight mixing induced by the quantum jumps. For $t>T/2$ the qubit is now predominantly in the eigenstate with greater loss, seeding non-Hermitian evolution toward the eigenstate $\ket{-}$. At the end of the parameter sweep, the qubit has undergone a switch between eigenstates, induced by the small perturbation of quantum jumps. This transition is accompanied by a sharp increase in the entropy as shown in Fig. \ref{fig4}(d).

Similar nonadiabatic state/energy transfer has been observed when dynamically encircling an EP \cite{Xu2016,Doppler2016,Choi2017,Zhang2018} as a result of nonadiabatic coupling between eigenstates and non-Hermitian gain-loss effects \cite{Milburn2015}. To verify that our parameter variation is sufficiently slow to prevent this nonadiabatic coupling, we plot the calculated dynamics in the absence of quantum jumps in Fig. \ref{fig4}(c), observing that there is no eigenstate switch. This reveals how quantum jumps effectively serve as a new source of nonadiabatic coupling, breaking adiabatic evolution even when parameter variation is sufficiently slow.  

Quantum jumps, even when introduced at very modest rates, produce significant effects on non-Hermitian dynamics. The dissipation induced by these jumps is greatly enhanced by proximity to the EP, with dynamics driven by non-Hermitian evolution. In addition, quantum jumps introduce a new timescale relevant to adiabatic state transport in non-Hermitian systems. Our study elucidates the role and effect of dissipation on quantum non-Hermitian evolution, highlighting how controlling these dissipation mechanisms will be critical for harnessing non-Hermiticity and complex energies in quantum information processing and quantum sensing \wcr{\cite{Wiersig2020,Pick2019,kumar2021, kumar2021_2, khandelwal2021, minganti2021}}. 


\begin{acknowledgments}
This research was supported by NSF Grant No. PHY-1752844 (CAREER), AFOSR MURI Grant No. FA9550-21-1-0202, and use of facilities at the Institute of Materials Science and Engineering at Washington University. 
\end{acknowledgments}



	\widetext
	\textcolor{white}{.}
	\newpage  
	
	\newpage
	\widetext
	\begin{center}
		\textbf{\large Supplemental Information}
	\end{center}
	
	In these supplementary materials, we \wcr{describe our experimental setup,} \krev{experimental sequences,} provide calculations of the \wcr{spectra of the hybrid-Liouvillian superoperators} in detail, and also discuss the decoherence effect led by quantum jumps.

\section{A. Experimental setup}

\wcr{Our setup consists of a transmon superconducting circuit dispersively coupled to a three-dimensional copper microwave cavity. The transmon is composed of a pair of Josephson junctions in a SQUID geometry shunted by a capacitor. It is fabricated through double-angle evaporation and oxidation of aluminium on silicon substrate. The transmon transition frequency can be adjusted by tuning a d.c. magnetic flux through the SQUID loop. \krev{The strongly coupled port of the cavity is connected via a microwave cable to an impedance mismatch, which creates a frequency dependence of the density of states of the electromagnetic field near the transition frequencies of the transmon. By flux tuning the transmon, different decay rates can be selected for the chosen transitions \cite{Naghiloo2019}}. In experiments, the transition frequencies are about $\omega_{ge}/2\pi=5.71\,\mathrm{GHz}$, $\omega_{ef}/2\pi=5.41\,\mathrm{GHz}$, with the corresponding charging energy $E_c/h = 270\,\mathrm{MHz}$ and the Josephson energy $E_J/h = 16.6\,\mathrm{GHz}$.}

\begin{figure}[hbt!]
    \includegraphics{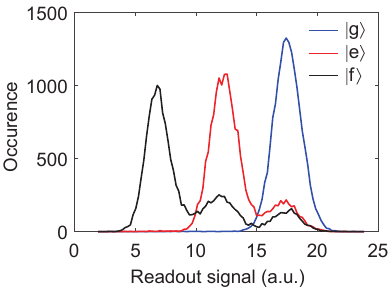}%
    \caption{\label{SIfig_readout} \wcr{Histogram of integrated single-shot readout signal, with the three states well separated.}}
\end{figure}

\wcr{The transmon circuit is dispersively coupled to the microwave cavity (with the dressed resonance frequency $\omega_c/2\pi = 6.684\,\mathrm{GHz}$ and decay rate $\kappa_c/2\pi = 5\,\mathrm{MHz}$) at a rate $g/2\pi= 65\,\mathrm{MHz}$, with the coupling rates for $|e\rangle$ and $|f\rangle$ levels given by $\chi_e/2\pi = -2\,\mathrm{MHz}$ and $\chi_f/2\pi = -11\,\mathrm{MHz}$. We achieve high-fidelity, single-shot readout of the transmon state in the energy eigenbasis by probing the microwave cavity with a weak signal and detecting its phase shift, assisted by a Josephson parametric amplifier with over $20\,\mathrm{dB}$ gain and instantaneous bandwidth of $50\,\mathrm{MHz}$. The readout fidelity for $|g\rangle$, $|e\rangle$, $|f\rangle$ are about $77\%$, $78\%$, $98\%$, respectively (Fig.~\ref{SIfig_readout}). In quantum state tomography, a $\pi/2$ pulse is used to rotate the qubit about $X$ and $Y$ axes, followed by readout in the energy eigenbasis. The duration of $\pi$ pulses used in the sequence are $\tau_{ge}=34\,\mathrm{ns}$ and $\tau_{ef}=30\,\mathrm{ns}$, respectively.  Each data set acquired with $10^4$ repetitions per time point, of this approximately 100-200 successful postselections contribute to the tomography. }

\section{B. Experimental sequences}

\krev{This section details the experimental sequences utilized for each experiment presented in the main text. }

\krev{Figure~\ref{fig2} displays the population dynamics versus time for different values of $J$ in the unbroken regime.  Experimentally, the transmon is assumed to start in the state $\ket{g}$ through thermalization with the cryogenic environment. Two resonant $\pi$ pulses transfer the state to $\ket{f}$. The pulse durations are $34$ ns and $30$ ns for the $\{\ket{g},\ket{e}\}$ and $\{\ket{e},\ket{f}\}$ manifolds, respectively. Once the system is prepared in $\ket{f}$, a resonant drive is applied to the $\{\ket{e},\ket{f}\}$ transition, creating the Hamiltonian term (in the rotating frame) $H_c = J(\ket{e} \bra{f} + \ket{f}\bra{e})$. The system evolves under this Hamiltonian for a variable duration of time before the coupling drive is abruptly turned off ($J=0$), and a high-fidelity, single-shot readout of the transmon state is applied as described above. Any experimental sequences where the circuit is found in state $\ket{g}$ are disregarded, allowing us to study the evolution of the population of the state $\ket{f}$ in the sub-ensemble. These populations are displayed versus time in  Fig.~\ref{fig2}b for different evolution times and three select values of $J$. From a larger set of data at different values of $J$, we extract the oscillation frequency and decay rate as displayed in Fig.~\ref{fig2}c.  }

\krev{Figure~\ref{fig3} displays the time evolution of the Bloch components versus time in the broken regime, where the system is prepared in the eigenstate of $H_\mathrm{eff}$ with greater loss. \wcrr{We first locate the eigenstate in the broken regime.} For this, the qubit is initialized in $\ket{f}$ (as for Fig.~\ref{fig2}), and an additional rotation is used to prepare} \wcrr{different superpositions of $\ket{e}$ and $\ket{f}$ on the $Y$--$Z$ plane of the Bloch sphere. After each preparation, a resonant coupling ($J=0.85\ \mathrm{rad}/\mu\mathrm{s}$) is suddenly applied. The superpostion state that exhibits the same $\ket{f}$ state population at $t=0$ and $t=300$ ns in the $\{\ket{e},\ket{f}\}$ submanifold is approximated as an eigenstate (i.e., it is a ``stationary state"). With this calibration, we proceed to study the time evolution of the Bloch components of the eigenstate. For this, the qubit is initialized by using the calibrated pulses, and then the resonant coupling is suddenly applied.} \krev{After a variable duration, the coupling is abruptly turned off and quantum state tomography is applied as described above. As before, any experimental sequences where the circuit is found in state $\ket{g}$ are disregarded.}

\begin{figure}[hbt!]
    \includegraphics{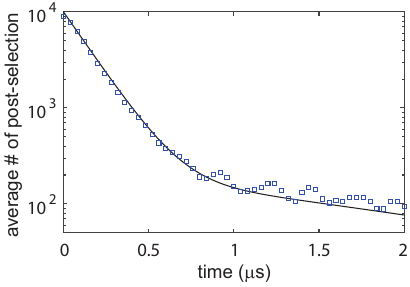}%
    \caption{\label{SIfig_post} \wcrr{The time evolution of average number of successful postselection in the $10^4$ experimental trials. The solid curve is a theoretical estimation from the Lindblad master equation.}}
\end{figure}

\krev{To estimate the statistical and technical error in the tomography measurements displayed in Fig.~\ref{fig3} we can examine the standard error of the mean due to the binomial distributed tomography results. With $10^4$ experimental trials, and $>100$ successful postselections at the longest times, the standard error is of order $0.05$, similar to the size of the data points. Rather, the observed fluctuations are dominated by stability of the experimental apparatus over long periods of time. Here, we repeat the measurement 10 times, and display the standard deviation of the experimental results as an error band in Fig.~\ref{fig3}.}
\wcrr{Figure~\ref{SIfig_post} shows the average number of successful postselections (i.e., when the transmon remains in the $\{\ket{e},\ket{f}\}$ submanifold) out of $10^4$ trials, where the two distinct decay rates also indicate a transition from the eigenstate with more loss to the one with less loss.}

\krev{Figure~\ref{fig4} displays data where the detuning ($\Delta$) of the coupling drive is tuned dynamically in time. The system is prepared in $\ket{f}$ as above, followed by a resonant $\pi/2$ rotation ($15$ ns) to prepare the state $\ket{+x}$, which is near the eigenstate of the effective Hamiltonian for $J=30\ \mathrm{rad}/\mu\mathrm{s}$. After preparing the state, the coupling is abruptly turned on, and tuned in real time. After a variable duration of evolution, and variable progression along the tuning, the coupling is abruptly turned off to perform quantum state tomography (as above). } \wcrr{The same method is used here to estimate the statistical and technical errors, and the standard deviation of the experimental results as an error band is provided in Fig.~\ref{fig4}.}

\section{C. Matrix representation and spectra of hybrid-Liouvillian superoperators}

The dynamics of the three-level quantum system in our study is described by a Lindblad master equation 
\begin{equation}
\frac{\partial \rho_{\mathrm{tot}}}{\partial t} = -i [H_\mathrm{c}, \rho_{\mathrm{tot}}] + \sum_{k=e,f} [ L_k \rho_{\mathrm{tot}} L_k^\dag - \frac{1}{2} \{L_k^\dag L_k, \rho_{\mathrm{tot}}\} ],
\end{equation}
where $\rho_{\mathrm{tot}}$ denotes a $3 \times 3$ density operator. The Lindblad dissipators $L_e = \sqrt{\gamma_e} |g\rangle \langle e|$ and $L_f = \sqrt{\gamma_f} |e\rangle \langle f|$ describe the energy decay from $|e \rangle$ to $|g \rangle$ and from $|f \rangle$ to $|e \rangle$, respectively. A microwave drive is applied to the \{$|e\rangle,|f\rangle$\} submanifold, and in the rotating frame $H_\mathrm{c} = J (|e \rangle \langle f| + |f \rangle \langle e|) + \Delta/2 (|e \rangle \langle e| - |f \rangle \langle f|)$, where $\Delta$ is the frequency detuning (relative to the $|e\rangle$--$|f\rangle$ transition) of the microwave drive that couples the states at rate $J$. 

In the absence of $L_f$, non-Hermitian evolution in the $\{\ket{e},\ket{f}\}$ submanifold can be isolated by eliminating quantum jumps from the $|e\rangle$ level. The resulting dynamics is governed by
\begin{equation}
\frac{\partial \rho}{\partial t} = -i [H_\mathrm{c}, \rho] 
- \frac{1}{2} \{L_e^\dag L_e, \rho\} \equiv \wc{\mathcal{L}}\rho,
\end{equation}
where $\rho$ denotes a $2 \times 2$ density operator and we define $H_\mathrm{eff} = H_\mathrm{c} -iL_e^\dag L_e/2$ and $\wc{\mathcal{L}} \rho = -i (H_\mathrm{eff} \rho - \rho H_\mathrm{eff}^{\dagger})$. 

To study the \wcr{hybrid-}Liouvilian spectra and exceptional points, we first represent the \wcr{hybrid-}Liouvillian superoperator in a matrix form \cite{Minganti2019,Minganti2020}, given by
\begin{equation}
    \wc{\mathcal{L}}^\mathrm{matrix} = -i(H_c\bigotimes I - I\bigotimes H_c^\mathrm{T}) 
    - \frac{L_e^{\dagger} L_e\bigotimes I}{2} - \frac{I \bigotimes L_e^\dagger L_e}{2},
\end{equation}
where $\bigotimes$ represents a Kronecker product operation and $\mathrm{T}$ represents the transpose. Accordingly, the density operator is written in a vector form,
\begin{equation}
    \rho = 
    \begin{pmatrix}
    \rho_{ee} & \rho_{ef} \\
    \rho_{fe} & \rho_{ff}
    \end{pmatrix}
    \rightarrow
    \begin{pmatrix}
    \rho_{ee}\\
    \rho_{ef}\\
    \rho_{fe}\\
    \rho_{ff}
    \end{pmatrix}.
\end{equation}
When the frequency detuning of microwave drive $\Delta=0$,
\begin{equation}
    \wc{\mathcal{L}}^\mathrm{matrix} = 
    \begin{pmatrix}
    -\gamma_e & iJ & -iJ & 0 \\
    iJ & -\gamma_e/2 & 0 & -iJ \\
    -iJ & 0 & -\gamma_e/2 & iJ \\
    0 & -iJ & iJ & 0
    \end{pmatrix},
\end{equation}
and accordingly,
\begin{equation}
    H_\mathrm{eff} = 
    \begin{pmatrix}
     - i\gamma_e/2 & J\\
    J & 0
    \end{pmatrix}.
\end{equation}

\begin{figure}[hbt!]
    \includegraphics{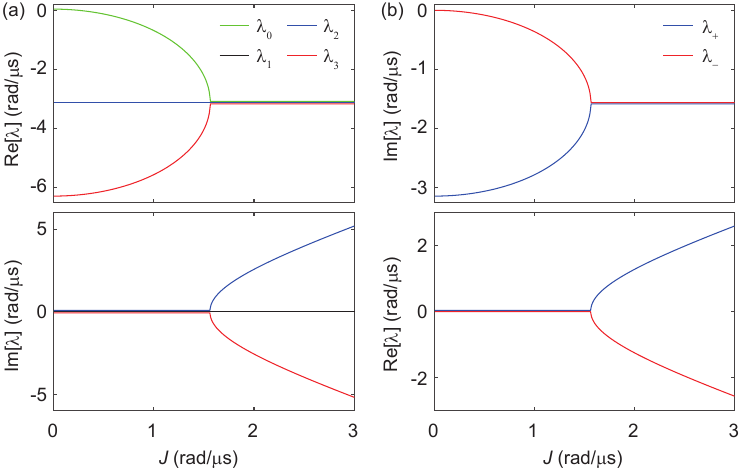}%
    \caption{\label{SIfig1} Eigenvalues of the \wcr{hybrid-}Liouvillian superoperator $\mathcal{L}_0$ (a) and the effective non-Hermitian Hamiltonian $H_\mathrm{eff}$ (b) at different drive amplitudes $J$. The parameters used are $\gamma_e = 6.25\,\mathrm{\mu s^{-1}}$, $\gamma_f = 0$, $\gamma_\phi = 0$, and $\Delta = 0$. The curves have been slightly offset for clarity.
    }
\end{figure}

The eigenvalues of $H_\mathrm{eff}$ and $\mathcal{L}^\mathrm{matrix}$ are provided in Fig.~\ref{SIfig1}. The \wcr{hybrid-}Liouvillian spectra is ordered as $\mathrm{Re}[\lambda_0] \geq \mathrm{Re}[\lambda_1] \geq \mathrm{Re}[\lambda_2] \geq \mathrm{Re}[\lambda_3]$. Due to the `$-i$' term in Eq. (1), the real and imaginary parts of \wcr{hybrid-}Liouvillian spectra should be compared to the imaginary and real parts of the spectra of $H_\mathrm{eff}$, respectively. 
Both spectra show an EP at $J=\gamma_e/4$.

\begin{figure}[hbt!]
    \includegraphics{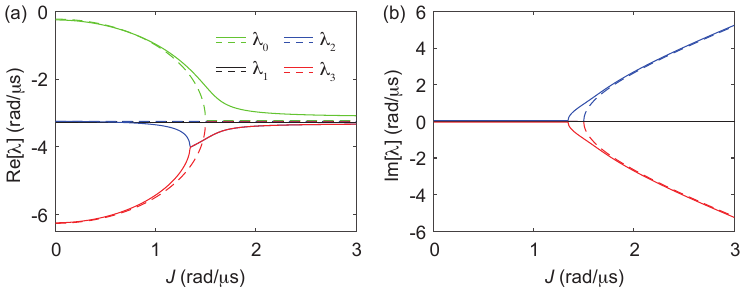}%
    \caption{\label{SIfig2} \wcr{Hybrid-}Liouvillian spectra $\lambda_{i=0,1,2,3}$ at different drive amplitudes $J$ with (solid curves) and without (dashed curves) quantum jumps from the energy decay of the $|f\rangle$ level. The parameters used are $\gamma_e = 6.25\,\mathrm{\mu s^{-1}}$, $\gamma_f = 0.25 \,\mathrm{\mu s^{-1}}$, $\gamma_\phi = 0$, and $\Delta = 0$. The jumps lift the third-order degeneracy creating a gap in the real part of the spectra, responsible for the decoherence enhancement near the EP. The remaining degeneracy (between $\lambda_2$ and $\lambda_3$) is shifted to a lower value of $J$. The curves have been slightly offset for clarity. }
\end{figure}

\begin{figure}[hbt!]
    \includegraphics{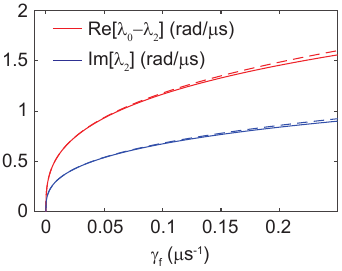}%
    \caption{\label{SIfig_cube_root} \wcr{Dependence of $\mathrm{Re}[\lambda_0-\lambda_2]$ (red solid curve) and $\mathrm{Im}[\lambda_2]$ (blue solid curve) on $\gamma_f$, where $J=(\gamma_e-\gamma_f)/4$ at each $\gamma_f$. The dashed red and blue curves are calculated from Eqs.~\ref{eqn_cube_root0} and \ref{eqn_cube_root1} that exhibit a cube-root dependence on $\gamma_f$. The parameters used are $\gamma_e = 6.25\,\mathrm{\mu s^{-1}}$, $\gamma_\phi = 0$, and $\Delta = 0$}.}
\end{figure}

We now consider the effect of the \wcr{jump operator} 
$L_f$. It affects the qubit dynamics in two aspects. First, it modifies the effective non-Hermitian 
Hamiltonian $H_\mathrm{eff}^\prime = H_\mathrm{eff} - iL_f^\dag L_f/2$, and the corresponding \wcr{hybrid} Liouvillian is denoted as $\mathcal{L}_0$ (See Table I). Second, the quantum jumps abruptly change the qubit state, the effect of which is described by the \wcr{hybrid-}Liouvillian superoperator $\mathcal{L}_1 \rho= L_f \rho L_f^\dagger$ which has no Hamiltonian counterpart. The matrix form of $\mathcal{L}_1$ can be calculated by using $\mathcal{L}_1^\mathrm{matrix} = L_f\bigotimes L_f$ 
(See Table I). Figure~\ref{SIfig2}  presents the \wcr{hybrid-}Liouvillian spectra with ($\mathcal{L}_0+\mathcal{L}_1$) and without ($\mathcal{L}_0$) considering quantum jumps. 
\wcr{The eigenvalues of $\mathcal{L}_0$ are given by
\begin{gather} 
\lambda_{0,1} = -(\gamma_e + \gamma_f)/2\\
\lambda_{2,3} = -(\gamma_e + \gamma_f)/2 \pm 1/2 \sqrt{(\gamma_e-\gamma_f)^2-16J^2}
\end{gather} 
With the effect of quantum jumps ($\mathcal{L}_1$), the eigenvalues of $\mathcal{L}_0 + \mathcal{L}_1$ are given by
\begin{gather} 
\lambda_{0} = -\frac{\gamma_e + \gamma_f}{2} - \frac{M}{6 N^{1/3}} + \frac{1}{6} N^{1/3}\\
\lambda_1 = -(\gamma_e + \gamma_f)/2\\
\lambda_{2,3} = -\frac{\gamma_e + \gamma_f}{2} + \frac{(1\pm i\sqrt{3})M + (-1\pm i\sqrt{3})N^{2/3}}{12 N^{1/3}}
\end{gather} 
with $M = 48J^2 - 3(\gamma_e-\gamma_f)^2$ and $N = 216 \gamma_f J^2 + \sqrt{(216 \gamma_f J^2)^2 + M^3}$.
}

\wcr{The quantum jumps ($\mathcal{L}_1$) as a perturbation lift the eigenvalue degeneracy at the third-order Liouvillian EP (i.e., $M=0$). On one hand, the splitting of the real part of eigenvalues leads to decoherence, determined by the eigenvalue difference $\mathrm{Re}[\lambda_0 - \lambda_{2,3}]$. On the other hand, the nonzero imaginary parts of $\lambda_{2,3}$ leads to oscillatory behavior. For small $\gamma_f$ limit,
\begin{gather}
\mathrm{Re}[\lambda_0 - \lambda_{2,3}] = \frac{3}{4} (\gamma_e^2 \gamma_f)^{1/3},
\label{eqn_cube_root0}\\
\mathrm{Im}[\lambda_{2,3}] = \frac{\sqrt{3}}{4}(\gamma_e^2 \gamma_f)^{1/3}.
\label{eqn_cube_root1}
\end{gather} Both effects are enhanced due to the cube-root topology of the hLEP (Fig. \ref{SIfig_cube_root}). 
}

\begin{figure}[hbt!]
    \includegraphics{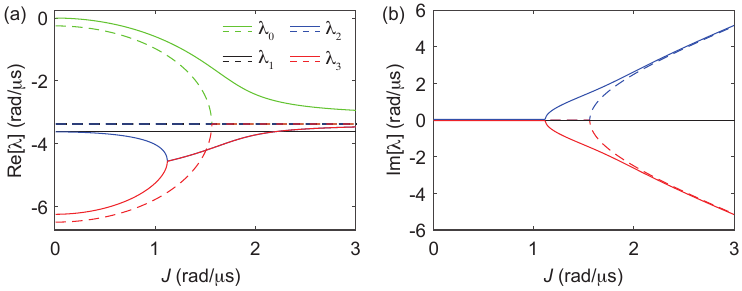}%
    \caption{\label{SIfig3} \wcr{Hybrid-}Liouvillian spectra $\lambda_{i=0,1,2,3}$ at different drive amplitudes $J$ with (solid curves) and without (dashed curves) quantum jumps from the pure dephasing of the $|e\rangle-|f\rangle$ submanifold. The parameters used are $\gamma_e = 6.25\,\mathrm{\mu s^{-1}}$, $\gamma_\phi = \wc{0.5} \,\mathrm{\mu s^{-1}}$, $\gamma_f = 0$, and $\Delta = 0$. The curves have been slightly offset for clarity.}
\end{figure}

Similarly, we can include the effect of pure dephasing in the submanifold, described by a jump operator $L_{\phi} = \sqrt{\gamma_{\phi}/2}\sigma_z$ 
with dephasing rate $\gamma_\phi$ (see Table I and Fig.~\ref{SIfig3}).

\begin{table}[hbt!]
 \begin{center}
 \begin{tabular}{|c|c|c|} 
 \hline
 dissipation & $\mathcal{L}_0$ & $\mathcal{L}_1$ \\ [0.5ex] 
 \hline
 $\gamma_f$ &  
$\begin{pmatrix}
-\gamma_e & iJ & -iJ & 0 \\
iJ & -(\gamma_e+\gamma_f)/2 & 0 & -iJ \\
-iJ & 0 & -(\gamma_e+\gamma_f)/2 & iJ \\
0 & -iJ & iJ & -\gamma_f
\end{pmatrix}$
 & $\begin{pmatrix}
0 & 0 & 0 & \gamma_f \\
0 & 0 & 0 & 0 \\
0 & 0 & 0 & 0 \\
0 & 0 & 0 & 0
\end{pmatrix}$ \\ 
 \hline
 $\gamma_\phi $ & $\begin{pmatrix}
-(\gamma_e+\gamma_\phi\wc{/2}) & iJ & -iJ & 0 \\
iJ & -(\gamma_e+ \gamma_\phi)/2 & 0 & -iJ \\
-iJ & 0 & -(\gamma_e+\gamma_\phi)/2 & iJ \\
0 & -iJ & iJ & -\gamma_\phi\wc{/2}
\end{pmatrix}$ &     
$\begin{pmatrix}
\gamma_\phi & 0 & 0 & 0 \\
0 & -\gamma_\phi & 0 & 0 \\
0 & 0 & -\gamma_\phi & 0 \\
0 & 0 & 0 & \gamma_\phi
\end{pmatrix}/2$  \\
 \hline
\end{tabular}
\end{center}
\caption{Matrix form of the \wcr{hybrid-}Liouvillian superoperators $\mathcal{L}_0$ and $\mathcal{L}_1$ under the dissipation of spontaneous emission of the $|f\rangle$ level at a rate $\gamma_f$ and the pure dephasing of the $|e\rangle-|f\rangle$ submanifold at a rate $\gamma_\phi$. 
}
\end{table}

\section{D. Decoherence in non-Hermitian dynamics with quantum jumps}

The $\mathcal{PT}$ symmetry breaking transition of a non-Hermitian qubit has been reported in Ref. \cite{Naghiloo2019}, manifested as a transition of population dynamics from exhibiting exponential decay in the broken regime to anharmonic sinusoidal oscillations in the unbroken regime. The quantum jumps within the qubit lead to decoherence, eventually resulting in a steady state. As shown in Fig.~\ref{fig2} of the main text, the decoherence effect manifests as a decaying oscillation in the unbroken regime. Here we provide the connection between this decay rate and the \wcr{hybrid-}Liouvillian spectra. 

The dynamics of the density matrix $\rho(t)$ (except at the EP) can be written as
\begin{equation}
    \rho(t) = \sum_{i=0,1,2,3} c_i(0)e^{\lambda_i t}\rho_i, \label{rho_t_liou}
\end{equation}
where $\lambda_i$ ($\rho_i$) is the eigenvalue (eigenvector) \wcr{of the hybrid-Liouvillian superoperator}, and $c_i(0)$ is determined by the initial state. Given the initial state $|f\rangle$, $c_1(0)=0$;  therefore the eigenvalue $\lambda_1$ does not affect the dynamics. The evolution for the population at each level $\{|e\rangle$, $|f\rangle\}$ can be obtained from $\rho(t)$, that is, $P_e(t) = \rho_{ee}(t)$ and $P_f(t)=\rho_{ff}(t)$. The population in the $\{|e\rangle$, $|f\rangle\}$ submanifold then can be calculated from $P_e^n = P_e/(P_e+P_f)$ and $P_f^n = P_f/(P_e+P_f)$.

In the absence of quantum jumps, the four eigenvalues have the same real part (corresponding to the decay rate) in the unbroken regime, subsequently leading to undamped oscillation for $P_e^n$ and $P_f^n$. The oscillation frequency is determined by the imaginary part of the eigenvalue $\lambda_2$ (equivalently $\lambda_3$, since $\mathrm{Im}[\lambda_2] =-\mathrm{Im}[\lambda_3]$). The quantum jumps lift the degeneracy
, and the eigenvalues no longer have the same real part (see Fig.~\ref{SIfig2}). Therefore, the components in Eq.~\ref{rho_t_liou} feature different decay rates: the eigenvector $\rho_0$ with the \wcr{smallest} decay rate corresponds to the \wcr{effective} steady state, and the relaxation rate to the \wcr{effective} steady state is determined by $\mathrm{Re}[\lambda_0-\lambda_{2}]$, which is enhanced near the \wcr{hLEP}. In our study, the experimental results are fit to an exponentially decaying sine function, which is a good approximation to the theoretical model.

\end{document}